\documentclass{emulateapj}
\usepackage{times}
\usepackage{graphicx}

\slugcomment{To Appear in ApJ Letters}

\shorttitle{2FGL J1653.6$-$0159}
\shortauthors{Romani et al.}

\begin{document}

\title{2FGL J1653.6$-$0159: A New Low in Evaporating Pulsar Binary Periods}

\author{Roger W. Romani\altaffilmark{1,2}, Alexei V. Filippenko\altaffilmark{3}, and
S. Bradley Cenko\altaffilmark{4,5}} 
\altaffiltext{1}{Department of Physics, Stanford University, Stanford, CA 94305-4060,
 USA; rwr@astro.stanford.edu}
\altaffiltext{2}{Visiting Astronomer, Kitt Peak National Observatory and Cerro Tololo
InterAmerican Observatory, National Optical Astronomy Observatory, which is
operated by the Association of Universities for Research in Astronomy (AURA) under
cooperative agreement with the National Science Foundation.}
\altaffiltext{3}{Department of Astronomy, University of California, Berkeley, CA 94720-3411, USA}
\altaffiltext{4}{Astrophysics Science Division, NASA Goddard Space Flight Center, MC 661, Greenbelt, MD 20771, USA}
\altaffiltext{5}{Joint Space-Science Institute, University of Maryland, College Park, MD, 20742, USA}


\begin{abstract}

	We have identified an optical binary with orbital period $P_b=4488$\,s 
as the probable counterpart of the {\it Fermi} source 2FGL~J1653.6$-$0159.
Although pulsations have not yet been detected, the source properties are consistent with an
evaporating millisecond pulsar binary; this $P_b=75$\,min is the record low for a spin-powered 
system. The heated side of the companion shows coherent radial-velocity variations, with 
amplitude $K=666.9\pm7.5$\,km\,s$^{-1}$ for a large 
mass function of $f(M)=1.60\pm0.05$\,M$_\odot$. This heating suggests
a pulsar luminosity $\sim 3\times10^{34}$\,erg\,s$^{-1}$. The colors and spectra
show an additional blue component dominating at binary minimum. Its origin is, at present, unclear. 
This system is similar to PSR~J1311$-$3430, with a low-mass 
H-depleted companion, a dense shrouding wind and, likely, a large pulsar mass.
\end{abstract}

\keywords{gamma rays: stars --- pulsars: general}

\section{Introduction}

Of 1873 0.1--100\,GeV sources in the second {\it Fermi} Large Area Telescope (LAT) catalog,
over 1170 had statistically reliable lower 
energy counterparts \citep{2FGL}. Most are blazars and spin-powered 
pulsars (radio-selected and Geminga-like $\gamma$-ray selected). \citet{r12}
noted that the identification completeness of the brightest 250 LAT sources
was even higher with (at that time) no more than six remaining unidentified and
showed that variability and GeV spectral curvature classification flagged 
all six as having pulsar-like properties. Three of these 
sources have since been identified as millisecond pulsar (MSP) binaries with strong winds:
J2339$-$0533, a ``redback'' (Romani \& Shaw 2011; Kong et al. 2012), J1311$-$3430,
a ``black widow'' (Romani 2012; Pletsch et al. 2013), and J1227$-$4859, a low-mass X-ray binary (LMXB)/MSP
transition object (Ray et al. 2014). Here we report evidence that a fourth member
of this set, 2FGL~J1653.6$-$0159 (hereafter J1653), is also a MSP-driven evaporating binary.


	J1653 is located at Galactic latitude $|b|=25^\circ$, with a total 
Galactic extinction $A_V =0.63$\,mag at this position \citep{sf11}.
With a $22.5\sigma$ detection significance in 2FGL, an energy flux
of $F_{0.1-100\,{\rm GeV}}=3.4\times10^{-11}\,{\rm erg\,cm^{-2}\,s^{-1}}$, a
``variability index'' value of 17, and a ``curvature significance'' of 5.3, this is a steady 
source with a substantial spectral cutoff: a prime pulsar candidate. It has been 
searched for $< 30$\,Hz $\gamma$-ray pulsations \citep[e.g.,][]{blind} and for radio pulses
to millisecond periods \citep{ray12,barret13}, with no detection. Thus, it is unlikely
to be an isolated young pulsar or a persistent radio-loud MSP. We describe here 
an optical campaign to identify and characterize a counterpart.

\section{Photometry and Orbital-Period Estimate}

	J1653 is well localized; we used the the Goodman High Throughput Spectrograph (GHTS)
at the 4.2\,m SOAR telescope on 2013 August 10--12 (UT dates are used throughout) to
image the $2^\prime$ radius LAT error ellipse and examine low energy counterparts.
Conditions varied from poor ($\sim1.5^{\prime\prime}$) to wretched ($>5^{\prime\prime}$), and 
so the GHTS data were binned $2\times2$.  Our strategy (as used to identify J1311 and J2339) 
was to obtain initial 300\,s $g^\prime r^\prime i^\prime$ frames for near-simultaneous colors
(Aug. 10), followed by single-color photometric sequences: $15\times 300$\,s in $g^\prime$ (Aug. 10),
$29\times 180$\,s in $r^\prime$ (Aug. 11), and $32\times300$\,s in $i^\prime$ (Aug. 12). We were
particularly interested in the eight {\it Chandra} X-ray sources in the revised error
ellipse (see Cheung et al. 2012). Four of these had optical counterparts in
our frames. The brightest X-ray source had a blue optical counterpart showing evident variability
during the first hour's observation and was thus the prime target of the 
light-curve study.
The USNO-B1.0 position of this $B=20.4$\,mag optical source (and likely LAT counterpart) 
is $\alpha=16^{\rm h}53^{\rm m}38.069^{\rm s}$, $\delta=-01^\circ 58' 36.71''$ (J2000.0). 
Although many frames had limiting magnitudes as small as $\sim23$, this star
was easily detected in all observations; we made sure to cover at least one maximum during
each night's photometry.

This counterpart was also observed in Gunn $r$ (180\,s) and 
narrow-band H$\alpha$ (600\,s) with the MiniMo camera at the 3.6\,m WIYN 
telescope on 2012 February 19. The narrow-band image did not yield a
reliable magnitude, but the $r$-band frame could be calibrated to
SDSS $r^\prime$, and matched well to the SOAR light curve at the phase
determined for the best-fit period (see below).

	We searched image archives for exposures that may have included J1653, finding
useful coverage in the Catalina Sky Survey (CSS;
http://www.lpl.arizona.edu/css/index.html; Drake et al. 2009). 
These data are shallow, unfiltered, and sparse, but they show
significant modulation at the orbital period.  Using the SOAR, WIYN, and CSS points, we searched 
the folded light curves for the minimum phase dispersion 
with the IRAF Phase Dispersion Minimum (PDM; Stellingwerf 1978) script.
This is $P_b= 0.05194469(+10,-08)$\,d, with epoch $T_{\rm ASC}=  56513.48078\pm0.00052$ (MJD). 
Aliases appear at $P_b= 0.05193768$\,d and $P_b=0.05195178$\,d, but these provided
less satisfactory light curves. The $P_b$ uncertainty estimate in the last two digits 
comes from the range of the main minimum with a PDM statistic $\theta$ lower than that of the
best sidelobe minimum.  The folded light curves (two periods) are shown in Figure 1.
This period, 4488\,s = 1.25\,h, is shorter than that of any known rotation-powered
pulsar binary.

	For pulsar-heated companions, we expect the counterpart to be hottest
(bluest) at maximum brightness. Guided by simple sinusoids, one sees
that while $g^\prime-r^\prime$ does decrease slightly at the peak ($\phi_B=0.75$),
$r^\prime-i^\prime \approx 0$ throughout the orbit. Also, the minimum is flat and the
system is bluest in this region; at the light curve minimum another component
dominates the light from unheated secondary.  The relatively poor photometry prevents 
any detection of inter-orbit variability.

\begin{figure}[t!!]
\vskip 7.9truecm
\includegraphics{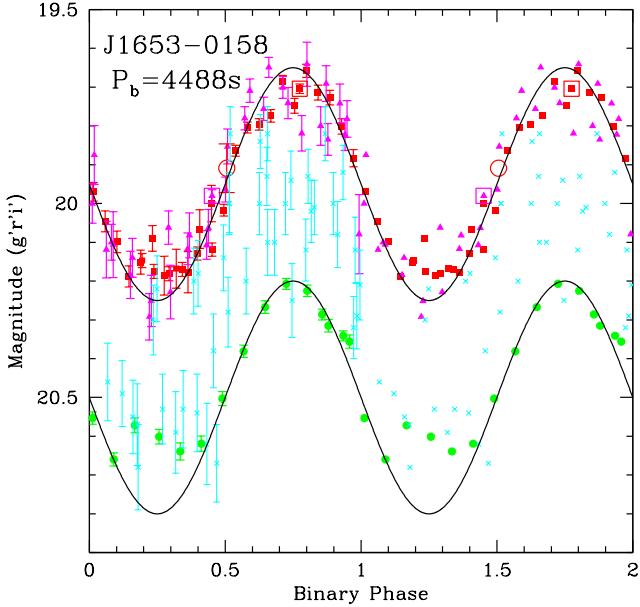}
\begin{center}
\caption{\label{LCfluxs} 
$g^\prime$(circles)/$r^\prime$(squares)/$i^\prime$(triangles) photometry of J1653$-$0158,
folded at the PDM-determined best-fit period.  The $r^\prime$ and $i^\prime$ points from 
2013 Aug. 10 are surrounded by open squares; the WIYN MiMo $r$ point is the open circle.
Two periods are shown, with error flags only during the first. CSS data (crosses) have 
uncertainties reduced by a factor of 2 for clarity. Simple sinusoids (with $g^\prime-r^\prime=0.55$\,mag)
are drawn to guide the eye.
}
\end{center}
\vskip -0.7truecm
\end{figure}

\section{Spectroscopy}

We attempted initial spectroscopy with the GHTS and 400\,line\,mm$^{-1}$ grating on 2013 August 11 
(MJD 56515), covering 3500--7000\,\AA\ at 7\,\AA\ resolution. With some of the best seeing 
of the run ($\sim1.3^{\prime\prime}$), we maintained reasonable throughput with
a fixed 1.07$^{\prime\prime}$ slit. Three 600\,s spectral exposures were obtained and
subjected to standard calibrations, using exposures of the spectrophotometric 
standard EG~274. These data exhibited no strong
emission lines, indicating that this was not a disk-dominated binary, and they showed a faint
secondary spectrum. Cross-correlation against stellar templates revealed velocity
shifts, but the low signal-to-noise ratio (S/N) and substantial velocity 
smearing gave large radial-velocity uncertainties.

	To explore the radial-velocity variations, we obtained two exposure
sequences with the Keck 10\,m telescopes. First, using the DEep Imaging Multi-Object Spectrograph 
(DEIMOS; Faber et al. 2003) on Keck-II (2013 September 10,
MJD 56545) we obtained $12\times180$\,s exposure covering 4450--9060\,\AA\ with
$\sim4.7\,$\AA\ resolution through a $1^{\prime\prime}$ slit. The short exposures minimized 
velocity smearing, and the sequence covered a full orbit. These data suggested a large 
radial-velocity variation, but had low S/N and lacked blue coverage. We also observed
with the Low Resolution Imaging Spectrometer (LRIS; Oke et al. 1995) on Keck-I on
2014 May 01 (MJD 56778), using the $1^{\prime\prime}$ long-slit and
the 5600\,\AA\ dichroic splitter. In the blue camera the 600/4000 
grism provided coverage to $<3400$\,\AA\ at $\sim4$\,\AA\ resolution, while the red camera 
employed the 400/8500 grating at $\sim7$\,\AA\ resolution. Red-side and blue-side exposures 
were 270\,s and 330\,s (respectively), and a sequence of seven exposures covered optical maximum. Observations of the
spectrophotometric standards HD~84937 and Feige~34 and the radial-velocity standard
HD~151288 were used in standard IRAF reductions, including optimal extraction. Again,
good evidence for radial-velocity variations was apparent, but here the spectra lacked full
orbital coverage.

	To complete our spectroscopic study, we conducted a second campaign with
Keck-I/LRIS, with the same spectroscopic configuration on 2014 May 24 (MJD 56802).
For both LRIS campaigns we set the slit at PA = +20$^\circ$ to include a star of comparable
brightness $29^{\prime\prime}$ south-southwest. Monitoring this G-type field star and the
night-sky lines allowed us to confirm the stability of the radial-velocity solution and
to connect photometry and velocity measurements between the observing runs. 
In this run we obtained a sequence of $43\times240$\,s
exposures with the red camera, binned a factor of 2 in the spatial direction to minimize
the readout time. The blue camera was run unbinned, and exposure times were adjusted
(250--285\,s, typically 265\,s) to keep the exposure midpoints approximately aligned with
those of the red camera. These observations covered three orbits; we also obtained exposures
of spectrophotometric standards and the G/K radial-velocity standards HD~122120,
HD~125184, and HD~125455.  

	A failure of the blue camera shutter early in the sequence
allowed counts to accrue during readout, adding up to $\sim7$\% continuum 
contamination at 5500\,\AA, less to the blue. Using a blue standard observed before
shutter failure and cross-calibrating with the field star (referenced to the 2014 
May 01 average spectrum), we developed a stable spectrophotometric solution. 
The red camera was not affected. Residual mis-match across the dichroic wavelength
suggests that the absolute spectrophotometry for $\lambda < 5550$\,\AA\, has an 
additional $\sim20$\% uncertainty beyond the usual variable slit losses. 

\subsection{Orbital Variations}

	Our study of the orbital variations starts with a comparison of spectra
at maximum brightness (``Day''; $0.6<\phi_B<0.9$; 14 exposures) and minimum 
brightness (``Night''; $0.1<\phi_B<0.4$; 13 exposures).  Figure 2 shows 
average spectra from these windows, Doppler shifted to the rest frame according 
to the best radial-velocity solution (see below). Although the true extinction is
not well known, we have dereddened using the full Galactic $A_V=0.63$\,mag, consistent 
with the X-ray spectrum (\S 4).
``Night'' is dominated by a nearly power-law component ($F_\lambda \propto 
\lambda^{-1.1\pm0.1}$); the step at $\sim5500$\,\AA\ likely represents residual 
flux-calibration uncertainties. With a lower $A_V=0.3$\,mag, the spectrum fits
somewhat less well as a power law with index $-0.75\pm0.2$.
The maximum-light ``Day'' spectrum includes a thermal excess (Diff), 
plotted along with main-sequence K5 ($T_{\rm eff} \approx 4400$\,K) and
F2 ($T_{\rm eff} \approx 6900$\,K) comparison spectra. The thermal excess is broader
than a stellar spectrum, evidently from the range of $T_{\rm eff}$ over the
heated ``Day'' face of the companion star. Low-S/N difference spectra at quadrature 
(``Dawn'' and ``Dusk'') give lower effective temperatures. The ``Dusk'' thermal excess
appears  brighter and hotter than that during ``Dawn,'' but this is presently of
low significance; such asymmetry can be explored using simultaneous multi-color photometry.

\begin{figure}[t!!]
\vskip 7.8truecm
\includegraphics{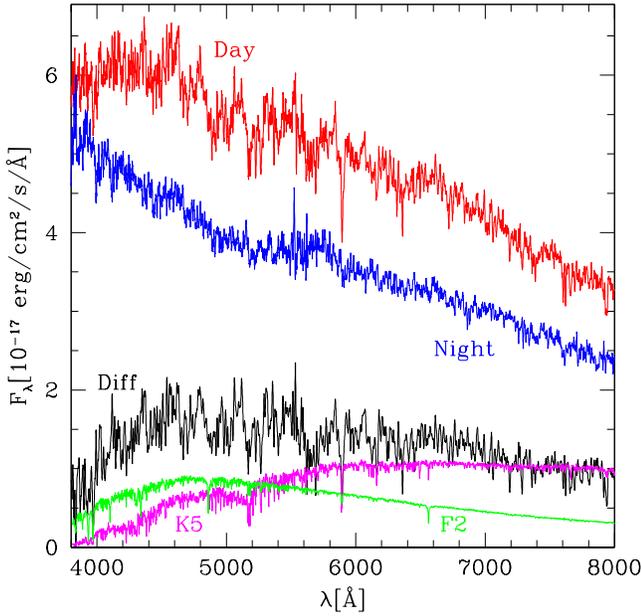}
\begin{center}
\caption{\label{DayNight} 
Keck-I/LRIS spectra of J1653$-$0158, showing maximum-brightness (``Day''; $0.6<\phi_B<0.9$) and 
minimum-brightness (``Night''; $0.1<\phi_B<0.4$) average spectra after dereddening and
shifting to the best-fit radial-velocity curve. The ``Day''-side thermal excess
is compared with two main-sequence stellar spectra.
}
\end{center}
\vskip -0.7truecm
\end{figure}

      The ``Night'' spectrum shows no spectral features; we find equivalent-width limits $<0.3$\,\AA\
for kinematic line width below $\sim10$\,\AA. The nature of this blue emission is unclear.
In contrast, the ``Day'' spectrum exhibits many broad and narrow absorptions matching
standard stellar spectral features. A number of the strongest metal lines can be 
individually identified in the ``Day'' average.
Hydrogen Balmer lines are not detected. Comparing with SDSS normal-star spectra, we find
equivalent-width limits of H$\alpha$, H$\beta$, and H$\gamma$ to respectively be $< 0.07$, 0.11, 0.14\,\AA\ 
in the ``Day'' spectrum versus 1.04, 0.66, 1.16\,\AA\ (K5~V) and 3.17, 4.15, 2.69\,\AA\ (F2~V). 
The equivalent widths should, however, be measured
for the thermal component of the spectrum; the limits for the lower-S/N difference 
spectrum are $< 0.4$\,\AA\ for all three species. Thus, H is weaker than in normal stars
by a factor of 2--10 in this Diff spectrum, depending on which $T_{\rm eff}$ dominates 
the ``Day''-side excess. While less impressive than for PSR J1311$-$3430, whose strongly
heated face shows a limit [H] $< 10^{-5}$ of normal, it appears that this
companion may have also lost appreciable hydrogen. We do not detect individual
He~I lines, but this is not surprising for the $T_{\rm eff}$ range seen here. Na~I $\lambda$5892 is,
as for PSR J1311$-$3430, very strong compared to the comparison stars.

\subsection{Orbital Radial Velocities}

	We can use the line features on the heated face 
to estimate the companion radial velocity. The measurements were made via 
cross-correlation, using the IRAF XCASO script \citep{km98}. We used the wavelength range 
3500--10,000\,\AA\ after excising regions near strong night-sky
lines and the immediate vicinity of the dichroic transition. The bulk of the correlation
significance stems from the metal line features at 4000--5500\,\AA, but significant
and consistent correlation signal was also measurable in the red portions of the spectra.
A range of stellar templates was used. While significant correlations and similar $v_r$
were obtained using early-M through F templates, the best correlations were obtained with 
K-type spectra.  We report here the results for a K5~V template.  Monitoring the
night-sky lines during each campaign and comparing the radial-velocity fits to
the comparison field star between Keck campaigns showed that our wavelength solutions
are stable to 0.1\,\AA\ $\approx5$\,km\,s$^{-1}$. Correlations against the
radial-velocity standards imply a similar accuracy for the absolute $v_r$. 

	The cross-correlation significance is monitored via the $R$ statistic. For
the May 24 data, the individual spectra near optical maximum had $R\approx7$--10 with 
radial-velocity accuracies $\sigma_{v_r}\approx15$--20\,km\,s$^{-1}$. Near quadrature
the accuracies dropped to $\sim30$\,km\,s$^{-1}$. Since ``Night''-side spectra lack
stellar features, the cross-correlation showed low significance with $R<2$. 
Although we do find similar ``Night'' velocities at a given phase from the
various observing runs, there is little systematic velocity trend and essentially
all the correlations are of low significance.  This suggests that the ``Night'' phase 
velocities, even when they show modest statistical errors, are generally not meaningful.

\begin{figure}[t!!]
\vskip 9.2truecm
\includegraphics{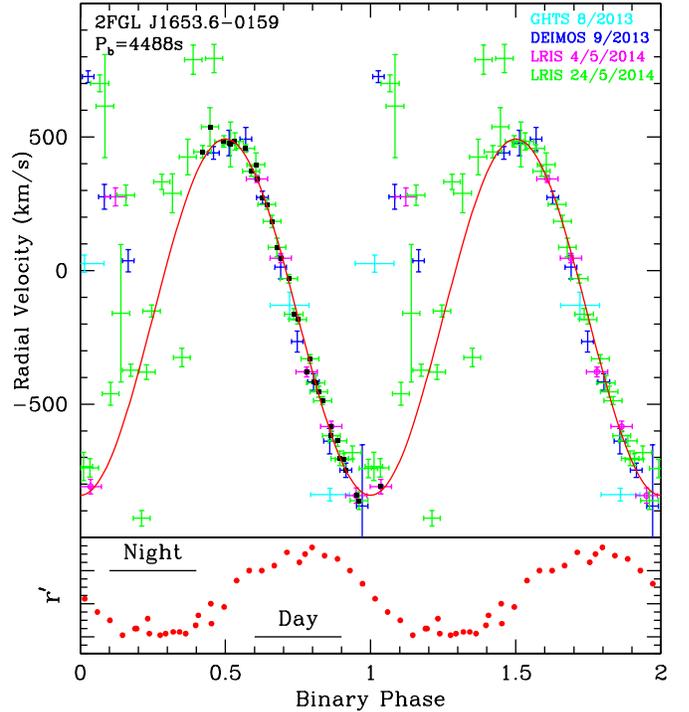}
\begin{center}
\caption{\label{RVfig} 
Radial velocities measured by cross-correlation of a K5~V template plotted on the photometric
ephemeris of \S 2. The $r^\prime$ light curve is plotted at bottom for reference and
two periods are shown.  Points with black dots are used to fit $K_2$ and $\Gamma$
for the simple sinusoid, shown by the line.
}
\end{center}
\vskip -0.7truecm
\end{figure}

	The cross-correlation velocities from our four spectroscopic campaigns are shown
in Figure 3, phased to the photometric ephemeris.  Over the phase range $0.4<\phi_B<1.1$, 
the velocity errors are small and, with a few outliers, the measurements from 
the various instruments are in good agreement.  
We performed a weighted least-squares fit to the measurements with $R>3$
in this phase range. Two 2014 May 24 LRIS points, at $\phi_B=0.94$ with $R=6.1$ and $\phi_B=0.99$ 
with $R=4.0$, lay several sigma from all best-fit curves. Excluding these points,
the fit uses the measurements marked with a black central dot in Figure 3 and 
results in a secondary velocity amplitude $K_2=666.9\pm7.5$\,km\,s$^{-1}$ and 
mean velocity $\Gamma=-174.6\pm5.1$\,km\,s$^{-1}$ with a fairly good $\chi^2$/DoF = 1.75.
Inclusion of the two dropped points decreases $K_2$ by 1.3$\sigma$ and $\chi^2$/DoF grows
to 2.52. Note that several low significance points in the range $0.9<\phi_B<1.1$
also lie $\sim70$\,km\,s$^{-1}$ redward of the best-fit curves. This, plus the
light-curve asymmetry and bluer spectrum at this phase, suggest some heating/wind activity
in the ``Dusk'' region, which may perturb the photospheric radial velocities. Similar,
but dramatically stronger effects are seen for J1311$-$3430 \citep{ret14}. 

	Our measured $0.4<\phi_B<1.1$ radial velocities follow the 
center of light of the secondary star, weighted by a K5~V spectrum. The resulting
mass function,
$$
f(M) = {{P K_2^3} \over {2\pi G}} = {{M_1^3\,{\rm sin^3}i}\over {(M_1+M_2)^2}} = 1.60 \pm 0.05\, {\rm M}_\odot,
$$
is a strict lower limit for the mass of the presumed pulsar. This 
value (the largest for any evaporating black-widow or redback-type pulsar) makes
it virtually certain that the unobserved, heat-producing object is a neutron star. 
It is unlikely that the blue spectrum seen at optical minimum is photospheric
emission from a white-dwarf primary, in accord with the lack of spectral features
and the sub-Rayleigh-Jeans spectral index. Since the
center of light of the heated face is inside the secondary center of mass
and since sin\,$i < 1$, the true primary mass is likely substantially higher. 

\section{Archival X-ray Data}

	Our counterpart, source 38 in the  21\,ks {\it Chandra X-ray Observatory} ACIS observation
of \citet{cet12}, provides $\sim350$ X-ray photons.
We fit these data to a power-law model with {\it CIAO/SHERPA}, finding a photon index $\Gamma=1.65_{-0.34}^{+0.39}$, 
absorption $N_{\rm H}=1.3_{-1.3}^{+1.8}\times10^{21}$\,cm$^{-2}$,
and unabsorbed flux $1.9_{-0.6}^{+0.9}\times10^{-13}\, {\rm erg\,cm^{-2}s^{-1}}$  
(0.5--8\,keV, $\chi^2_G$ statistic, 90\% uncertainty), in agreement with the \citet{cet12} measurements.
Note that $A_V =0.63$\,mag corresponds to $N_{\rm H}\approx1.2\times10^{21}$\,cm$^{-2}$,
so the X-rays suggest that J1653 is seen through much of the Galactic dust layer. Improved
measurements are needed to pin down the extinction and distance.  We folded the X-ray counts on the
photometric ephemeris and find no evidence for an eclipse or for any orbital variation,
although given the limited counts, shallow modulation could have been missed.
A similar fold of the LAT GeV photons also shows no evidence for orbital modulation.

\section{System Modeling and Conclusions}

	Lacking kinematic constraints from $\gamma$-ray or radio timing,
our picture of J1653 is still imprecise.
Some basic pulsar scaling laws can, however, be used to check consistency with
the evaporating binary scenario. First, $\gamma$-ray pulsars have a heuristic
luminosity $L_{\gamma,{\rm heu}} \approx ({\dot E}\, \times \, 10^{33}\, {\rm erg}\, {\rm s}^{-1})^{1/2} $
\citep{psrcat}. Next, X-ray emission from rotation-powered pulsars scales as
$L_X \approx 10^{-3}\,{\dot E}$ \citep{beck09}. The observed fluxes
$f_\gamma$ and $f_{\rm X, unabs}$ 
then give estimates of ${\dot E} \approx 3\times 10^{34}\, {\rm erg\, s^{-1}}$ and
$d\approx1.1$\,kpc.  \citet{bet13}
describe how pulsar spin-down flux (assumed isotropic at separation $a$) heats the companion
from a ``Night''-side $T_{\rm N}$ to a maximum $T_{\rm D}$ according to
$$
\eta {\dot E} \approx 4\pi a^2 \sigma (T_{\rm D}^4 - T_{\rm N}^4),
$$
finding a typical efficiency $\eta \approx 0.15$ for black-widow systems.
For J1653, we have $a\approx K_2 P_B/2\pi \approx 0.7\, {\rm R}_\odot$ and $T_{\rm D} >> T_{\rm N}$, so we can expect 
$T_{\rm D} \approx 5500\,{\rm K} (\eta_{0.15} {\dot E}_{34})^{1/4}$. The effective $T_{\rm D}$
is not well measured, but from our spectroscopy it seems lower than 6500\,K.
This implies $\eta_{0.15} {\dot E}_{34}\approx2$, somewhat lower than the $f_\gamma/f_{\rm X}$
estimate above, and suggesting a smaller $d\approx0.8$\,kpc as well.
Finally, the extinction-corrected optical flux from the heated face (Diff spectrum)
is $f_{\rm opt} \approx f_{-13} 10^{-13}\, {\rm erg\, cm^{-2} s^{-1}}$ with $f_{-13}\approx0.9$
using the bolometric correction for 6000\,K. From this we derive an effective companion radius of 
$R_\ast \approx  (4 f_{\rm opt}/\sigma)^{1/2} d/T_{\rm eff}^2 
\approx 0.10 f_{-13}^{1/2} d_{\rm kpc}/(T/6000\,{\rm K})^2 R_\odot$.  
This gives $R_\ast /a \approx 0.13\, d_{\rm kpc}/T_{6000}^2$,
which can be compared with the Roche radius $R_{\rm L2} /a \approx 0.46\, q^{-1/3}$.  
For a redback-type mass ratio $q\approx10$ this gives $R_{\rm L2}/a \approx0.2$, so the companion
would be well inside the Roche lobe. For a black-widow-type mass ratio $q \approx 100$,
$R_{\rm L2}/a \approx0.1$, so the companion fills the Roche lobe, even for somewhat larger
$T_{\rm eff}$, smaller $A_V$, and smaller $d_{\rm kpc}$.

	We would like to improve the mass and fill-factor estimates. 
The absence of
X-ray or $\gamma$-ray eclipses implies $i < 85^\circ$, so the minimum
pulsar mass ($K_2$ lower limit) is 1.62\,M$_\odot$ for a substellar 
(black-widow-type) companion. For a redback-type $\sim 0.2$\,M$_\odot$ companion, the
limit is $M_{\rm PSR} > 1.92\,$M$_\odot$. One generally
refines the constraints on the center-of-light to center-of-mass correction
factor $K_{\rm cor} = K_{\rm CoM}/K_{\rm CoL}$ and the inclination $i$ via light-curve
modeling. This is important since $K_{\rm cor}>1$ and
the pulsar mass grows as $M_{\rm PSR} \propto (K_{\rm cor}/{\rm sin}\,i)^3$. 
At present such modeling is not very constraining since our photometry is poor 
and nonsimultaneous.  Additional observations can, of course, remedy this. 

	More important, though, are the apparently nonthermal flux 
which dominates the optical light at minimum brightness and the possible 
light-curve asymmetries near maximum brightness. These suggest that a strong
evaporating wind distorts the light curve and makes it difficult to detect the
companion at minimum brightness. Simple heating models may
not be adequate. This also appears true for other black-widow systems with 
well-studied light curves, J1311 \citep{ret14} and PSR J1544+4937 \citep{tet14}.
Improved understanding of evaporative winds and their effect on the
light curves seem crucial for refined measurements of the short-period
black widows. J1653 certainly motivates such efforts: the small $P_b$, 
peculiar companion composition, and likely large $M_{\rm PSR}$ all flag this as 
an extreme member of the evaporating-pulsar population. Once $\gamma$ 
and/or radio pulses are detected, we plan to pursue such studies.
\bigskip

We thank Sasha Brownsberger and Matt Stadnik for assistance with the
photometric observations, and Melissa Graham, Patrick L. Kelly, and
WeiKang Zheng for assistance with the spectroscopic observations.
This work was supported in part by NASA grant NNX11AO44G.  A.V.F. was
supported by the Richard \& Rhoda Goldman Fund, the Christopher
R. Redlich Fund, the TABASGO Foundation, and NSF grant AST-1211916.
This research is based in part on observations obtained at the
Southern Astrophysical Research (SOAR) telescope, which is a joint
project of the Minist\'{e}rio da Ci\^{e}ncia, Tecnologia, e
Inova\c{c}\~{a}o (MCTI) da Rep\'{u}blica Federativa do Brasil, the
U.S. National Optical Astronomy Observatory (NOAO), the University of
North Carolina at Chapel Hill (UNC), and Michigan State University
(MSU).  Some of the data presented herein were obtained at the
W. M. Keck Observatory, which is operated as a scientific partnership
among the California Institute of Technology, the University of
California, and NASA; the Observatory was made possible by the
generous financial support of the W. M. Keck Foundation.

\end{document}